# Winter Road Surface Condition Recognition Using a Pre-trained Deep Convolutional Neural Network

**Guangyuan Pan**
Postdoc Fellow, Department of Civil & Environmental Engineering,
University of Waterloo,
Waterloo, ON, N2L 3G1, Canada
Email: g5pan@uwaterloo.ca

**Liping Fu***
Professor, Department of Civil & Environmental Engineering, University of Waterloo, Waterloo, ON, Canada,
N2L 3G1; Intelligent Transportation Systems Research Center, Wuhan University of Technology, Mailbox 125,
No. 1040 Heping Road, Wuhan, Hubei 430063
Email: lfu@uwaterloo.ca

**Ruifan Yu**
Master Student, David R. Cheriton School of Computer Science,
University of Waterloo,
Waterloo, ON, N2L 3G1, Canada
Email: ruifan.yu@uwaterloo.ca

**Matthew Muresan**
Ph. D. Student, Department of Civil & Environmental Engineering,
University of Waterloo,
Waterloo, ON, N2L 3G1, Canada
Email: mimuresa@uwaterloo.ca

## ABSTRACT

This paper investigates the application of the latest machine learning technique – deep neural networks for classifying road surface conditions (RSC) based on images from smartphones. Traditional machine learning techniques such as support vector machine (SVM) and random forests (RF) have been attempted in literature; however, their classification performance has been less than desirable due to challenges associated with image noises caused by sunlight glare and residual salts. A deep learning model based on convolutional neural network (CNN) is proposed and evaluated for its potential to address these challenges for improved classification accuracy. In the proposed approach we introduce the idea of applying an existing CNN model that has been pre-trained using millions of images with proven high recognition accuracy. The model is extended with two additional fully-connected layers of neurons for learning the specific features of the RSC images. The whole model is then trained with a low learning rate for fine-tuning by using a small set of RSC images. Results show that the proposed model has the highest classification performance in comparison to the traditional machine learning techniques. The testing accuracy with different training dataset sizes is also analyzed, showing the potential of achieving much higher accuracy with a larger training dataset.

## INTRODUCTION

Winter road surface condition (RSC) monitoring is of critical importance for winter road maintenance contractors and the traveling public. Real-time reliable RSC data can enable winter maintenance personnel to deploy the right type of maintenance treatments with the right amount of deicing materials at the right time, leading to significant savings in costs and reduction in salts. The traveling public could make more informative decisions on whether or not to travel, what mode to use and which route to drive for safer mobility. RSC monitoring is traditionally done by manual patrolling by highway agencies and maintenance contractors or using road weather information system (RWIS) stations (1-2). Manual patrolling provides very high spatial resolution with additional qualitative details, but suffering from the drawbacks of being subjective, labor-intensive, and time-consuming. On the other hand, RWIS stations benefit from providing continuous information on a wide range of road and weather conditions; however, they are costly and can only be installed a limited number of locations, limiting their spatial coverage. In recent years, new technologies have been developed to automate the RSC monitoring process, such as CCTV cameras, in-vehicle video recorders, smartphone-based system, and high-end imaging systems (3-7). However, these solutions have been found to be still limited in terms of working conditions and classification accuracy.

This research focuses on exploring the potential of applying one of the most successful machine learning models - deep learning for classifying winter road surface conditions based on image data. In Section 1, this paper first describes the idea behind the proposed RSC monitoring system and subsequently evaluates its performance as intended in the field. Section 2 reviews the previous studies that are related to this research, including image recognition and deep learning. Section 3 introduces the model that we propose. Section 4 describes the data information, designs the experiment, discusses model setting, and analyzes the results. The paper then concludes in Section 5.

## LITERATURE REVIEW

The image recognition problem has been studied extensively in the literature. Many machine learning models have been proposed in the past decades. In our previous research, three machine learning techniques, artificial neural networks (ANN), random trees (RT) and random forests (RF) were evaluated using the images of bare, partly snow-covered or fully snow-covered winter roads collected during the winter season of 2014. ANN, which is a black-box model, is commonly used to recognize patterns and model complex relationships among variables (8-9). Random Forest (RF) is an ensemble of classification trees that produces a result based on the majority output from the individual trees, in which each tree is constructed using a bootstrapped sample of the total data set (random sampling with replacement) (10). With a sufficient number of trees the predictions tend to converge, resulting in a reliable algorithm that is relatively robust to outliers and noise.

Deep learning (DL), or deep neural network (DNN), is a novel machine learning technique that has been widely explored and successfully applied for a variety of problems such as applications in image and voice recognition and games (11-12). In particular, convolutional neural networks (CNN) have recently shown great success in pattern recognition problems, such as large-scale image and video analysis. This achievement results from both the large public image repositories (such as ImageNet) and high-performance computing systems like GPU and the recent tensor processing unit (TPU) manufactured by Google (13-14). Since CNN is becoming very common in many machine learning fields, and better performances have been achieved by improving the original architecture and algorithms. For example, researchers have been proposing models with larger layers and deeper structures; however, the deeper the network is, the more difficult the training process is (15-16). In Karen Simonyan's work, the model, VGG16, secured the first and the second places in the localization and classification tracks respectively in ImageNet Challenge 2014 (17). However, deep learning is generally a big data technology, thus, in studies which contain insufficient samples; deep learning will struggle to learn useful features from the input. In these cases, the raw images often need to be preprocessed like other traditional machine learning approaches would do. In our research, we will present a simple yet effective method that builds a powerful image classifier, using raw data from only a small set of training examples. Our model is based on the model of VGG16 (17), which is pre-trained with learned features that are useful for most image recognition problems.

**DEEP LEARNING MODEL**

Convolutional neural network (CNN) is one of the deep learning models that have been especially successful for image classification. An example structure of CNN is shown in Figure 1(a), which is also included as one of the model options in our subsequent analysis.

*Pre-trained deep CNN Model Structure*

Instead of training a completely new CNN model, which often requires a significant amount of data and computational time, an alternative approach would be to make use of a CNN model which has already been trained with proven performance (19-20). Such a model would have already learned features that are useful for most computer vision problems; leveraging such features would allow us to reach a better accuracy than any method that would only rely on the available data. In this research, we use a pre-trained deep CNN model called VGG16 as has been introduced before.

Figure 1(b) shows the overall structure of the VGG16 deep CNN. The network can be divided into five convolutional blocks and a fully connected block. The first two convolutional blocks contain two convolutional layers with a receptive field with dimensions of $3 \times 3$ and convolutional kernels 64 and 128 respectively. The receptive field dimensions in this case refer to how big an area (pixels) the next layer can observe from the previous layer. The number of convolutional kernels (have carefully been studied in their paper) can decide how many features a convolutional layer can learn from its previous layer. The last three convolutional blocks contain three convolutional layers in each block; they also have a receptive field with dimensions of $3 \times 3$, and the convolutional kernels are 256, 512 and 512, respectively. The original raw images are first resized into a three channeled (Red, Green, Blue) image with dimensions of $150 \times 150$. The RGB values are then normalized by subtracting the RGB value of each pixel by the mean RGB value of the image. While this process results in some minor information loss, it significantly helps reduce computation times.

During training, the image is passed through a stack of convolutional layers, where only a very small receptive field of $3 \times 3$ is used. The convolution step is fixed to 1 pixel. Spatial pooling is applied to the output at the end of each of the five convolutional blocks. This is a technique commonly used in computer vision which applies a statistical measure across a group of pixels by scanning a window of a certain size across the input, similar to the process used in convolutional layers. In this model, max-pooling is used over a 2x2 window. The stack of convolutional blocks is followed by two fully connected (FC) layers after transforming the data to one dimension (which is called Flatten and Dense in Figure 1(b)). The two fully connected layers have 512 and 256 nodes respectively. The final layer is a softmax layer which gives classification results as a probability value for each class For the case of our problem of classifying RSC into one of three classes - barely, partly and fully coverage, the output layer will contain 3 nodes, one for each class, and the values in each of the nodes will correspond to the probability that the input image belongs to that particular category. The configuration of the fully connected layers is the same in all networks. The nodes in all hidden layers are rectifier linear units (ReLU), with detailed algorithms given in next section.

VGG16 has already learned features that are common to most image recognition problems. Thus, it is possible that merely recording the softmax predictions of the model over our data rather than recording all the features learned in convolutional layers would be enough to solve our classification problem. However, since the winter road condition images and classes are not included in the image database (ImageNet) that was used to train the VGG16 model, it is necessary to further train or fine tune this model with our domain specific data (RSC images), as discussed in the following section.

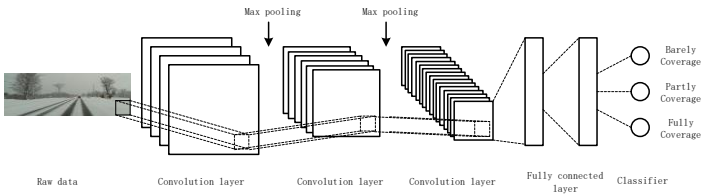

(a) Traditional CNN for image classification

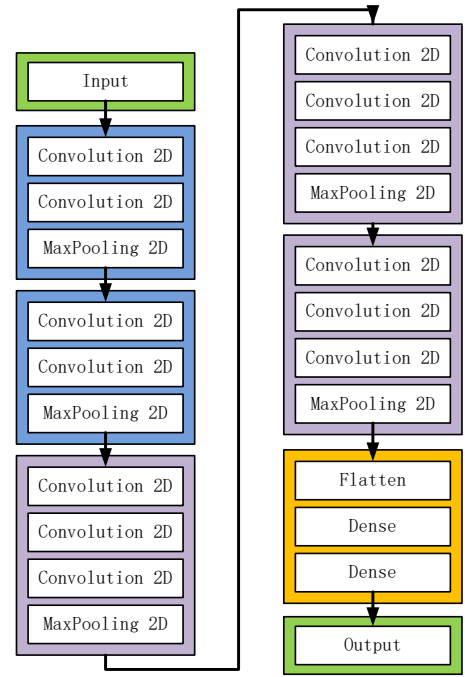

(b) Pre-trained deep CNN Model Structure
**FIGURE 1 CNN model**

*Pre-trained CNN Model Training and Fine-tuning Steps*

The strategy associated with this model is as follows.

1) Instantiate the convolutional part of the model - everything up to the fully-connected layers and then run this model on the training and validation data once to obtain the output. The convolutional function (Equation 1) and the ReLU transfer function (Equation 2) are used in this step. After computing each block (Figure 2), a Max pooling layer is used to reduce the dimensions of the output, using Equation (3).

$$x^l = f(w^{l-1} \cdot x^{l-1} + b^{l-1}) \quad (1)$$

$$f(x) = \max(0, x) \quad (2)$$

$$x = \max(matrix(i, j)) \quad (3)$$

In Equation (1), $x^{l-1}$ is the input of a convolutional layer $l$, $x^l$ its output, w and b the weight matrix and bias. In Equation (2), f(x) is the ReLU transfer function, which is the most popular function used in recent research. In Equation (3), matrix(i, j) is the input matrix to the max-pool layer at each step, and i and j are the dimensions (2 × 2 in this case).

2) Freeze all the convolutional layers and Max pooling layers from step 1, and train a small fully connected model on top of the model. The weights are updated using stochastic gradient decent (SGD). For computational reasons, we only store the features offline rather than adding our fully connected model directly on top of a frozen convolutional base and running the whole network. Running the model from a randomly initialed state is computationally expensive, especially if training is only done on the CPU.
3) To further improve the model, the last several convolutional blocks of the pre-trained model alongside the top-level classifier need to be fine-tuned. Fine-tuning is done on the fully trained network by re-training it on an additional training dataset using very small weight updates. To do this, we freeze the first 6 layers of the model (the first two blocks) and update the last three by a gradient decent method. The code we use is from an open source (23).

Note that in order to perform fine-tuning, all layers should start with properly trained weights. For instance, a randomly initialized fully-connected network on top of a pre-trained convolutional base, such as the case after step (1), should not be used. This is because the large gradient updates triggered by the randomly initialized weights would wreck the learned weights in the convolutional base. This is why we first train the top-level classifier, and only then start fine-tuning convolutional weights alongside it. We choose to only fine-tune the last three convolutional blocks rather than the entire network in order to prevent overfitting, since the entire network would have a very large entropic capacity and thus a strong tendency to over fit. The features learned by low-level convolutional blocks are more general and less abstract than those in higher level layers; consequently, it is sensible to freeze the first few blocks, which contain more general features, and only fine-tune the last several blocks (see more in Model Sensitivity). Fine-tuning should be done with a very slow learning rate, and typically with the stochastic gradient decent optimizer, which will ensure that the magnitude of the updates stays very small, so as not to wreck the previously learned features.

## EXPERIMENT

*Data Description*

The data used in this paper was collected from two highway sections in South-Western Ontario, Canada near Mount Forest as shown in Figure 2. It shows the primary test section on Highway 6, a two-lane, two-way, asphalt surfaced Class 2 highway, approximately 70km in length, with a winter average daily traffic (WADT) volume of 4900. The site has uniform geometrical features and few horizontal curves. This area experiences an annual average of 59 days of snowfall of at least 0.2cm (21). Data from this test site were captured during the winter of 2014.

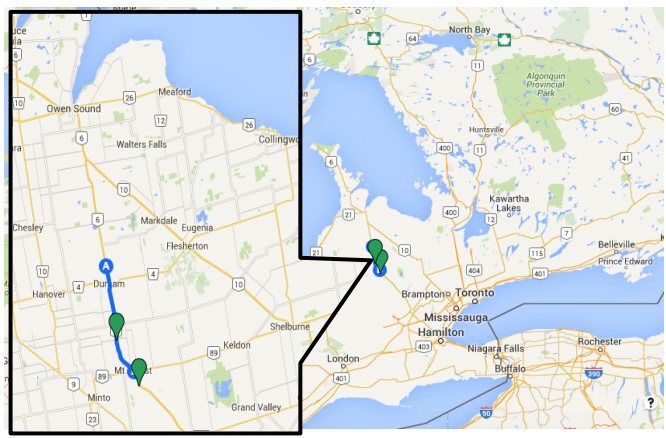

**FIGURE 2 Data collection location**

Images from the patrol vehicles and mobile data collection unit (MDCU) are captured at intervals of 350m and 450m, respectively. Each image is manually classified (ground truth) according to the five-class, three-class and two-class descriptions illustrated in Table 1, and the automatic image processing is designed to classify each image according to the descriptions given there. The five-class description categorizes RSCs by varying degrees of lateral snow coverage in accordance with the Transportation Association of Canada's RSCs description terminology, which is commonly used by winter maintenance personnel across Canada. The three-class description categorizes RSCs in accordance with Transportation Association of Canada's route reporting terminology, which is typically used to convey RSC information about maintenance routes to the general public.

**TABLE 1 Definition of Different Types of Lateral Snow Coverage**

| Sample Image | Description | Five-Class Description | Three-Class Description | Two-Class Description |
|---|---|---|---|---|
| 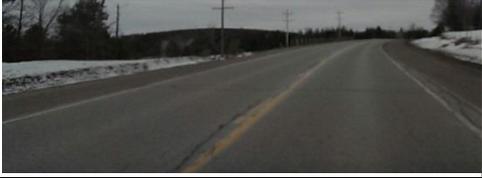 | At least 3 meters of the pavement cross-section in all lanes clear of snow or ice. | Bare | Bare | Bare |
| 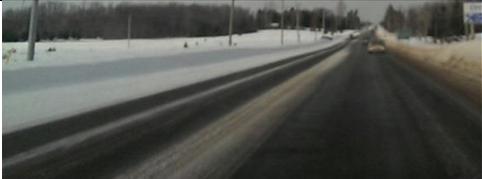 | Track between two wheel paths clear of snow or ice. | <25 (Essentially Bare) | Partly Snow Covered | With snow covered |
| 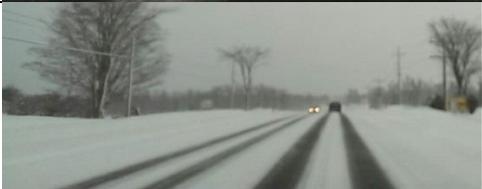 | Both wheel paths clear of snow or ice. | 25 to 50 | | |
| 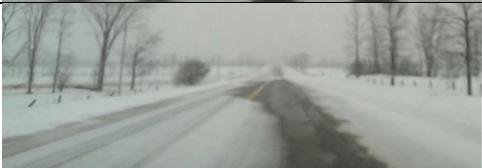 | Only one wheel path is clear of snow or ice. | 50 to 75 | | |
| 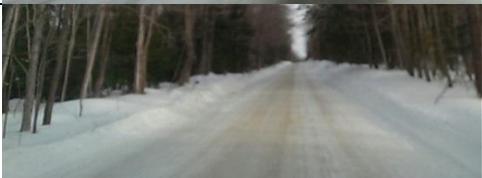 | No wheel path clear of snow or ice. | Fully Snow Covered | Fully Snow Covered | |

*Model Settings*

For pre-trained deep CNN models, a number of key configuration parameters can affect the accuracy and efficiency of the model. A sensitivity analysis is therefore necessary to determine the effect of key parameters such as the network structure used, learning rate, and the number of blocks fine-tuned.

As the convolutional layers in the model are pre-trained, the structure parameters that we configure in this model are defined by the number of fully connected layers and number of units (or nodes) in each layer. These settings could have a significant effect on the performance of the model. Too many layers and units can increase the computation and training time and cause the model to be overfitting to the training dataset, while too few layers and units may lead to poor feature learning and under-fitting. The significance of these parameters on the model performance means that these values must be chosen carefully. In paper (17), for the task ImageNet dataset with 1000 categories, the structure is set to be two layers with 4096 nodes in each one. As a result, a structure with two layers was chosen in this study (the last block as shown in Figure 1(b)). This structure strikes a balance between the need for a more powerful model and the desire to keep computation times lower. It offers more predictive power than a single layer model, but is significantly less time-consuming than a three layer model.

Besides the model structure, the selection of an appropriate learning rate is also very important. For the proposed model, learning rates for both the pre-training stage and fine-tuning stage must be specified, as they control how the weights of the connections between layers in each epoch (update weights for one time after training the whole training set) are updated. If the learning rate is too large, the weights may change too dramatically; on the other hand, if learning rate is too small, more epochs are needed, which not only increases the learning time but may also limit its chance to find the globally optimal solutions. In this study, the pre-training learning rate is set to be 0.001, and the fine-tuning learning rate is 0.0005.

In addition to the learning rate and model structure, the number of blocks that should be fine-tuned is also important to the performance. Intuitively, including more blocks in the fine-tuning process should increase the accuracy, as in the pre-trained deep CNN model, no road surface condition features are used in the original training. However, because the feature training in lower layers is similar, if the entire network is fine-tuned, it is possible that the feature identification capacity of the model may be reduced as the fine-tuning process may undo some of the learning progress made in the pre-training stage. Additionally, fine-tuning more blocks is more time consuming and computationally expensive. For all the models developed, the fine-tuning blocks are restricted to the last three, and the weights in the first two blocks and two max pooling layers are held constant.

*Sensitivity Analysis*

Since there are so many parameters need to train, for example, in the case of two-class description, the input is shrinking from 150x150 after each successive layer, until finally it is 4x4, however, the weights need to be updated are exploded from 1,792 to 2,359,808 in each layer. Thus, the more blocks we freeze in fine tuning, the faster. In this section, discusses the results of a sensitivity analysis conducted on the model settings described previously. For all the three experiments, the parameters settings follow the same procedure. First, during model pre-training, fully connected layers with dimensions 256-256-3, 512-256-3, 1024-512-3, and 2048-2048-3 are tested, as shown in Figure 3(a). After determining the optimal structure, pre-training learning rates ranging from 0.0001 to 0.01 and fine-tuning learning rates ranging from 0.0001 to 0.001 are tested, as shown in Figure 3(b) and 3(c) respectively. Finally the VGG16 the effect of freezing blocks 1 to 4 during fine-tuning is explored. The results are shown in Figure 3(d).

As we can see, with the increasing of the first fully connected layer, the performance increases. However, after researching a peak, the accuracy will not increase, and even drop down a little. This is probably because of over fitting, as our training samples are few. And pre-training learning rate 0.001 and fine-tuning learning rate 0.0005 give the best testing accuracy. In the last step, when only fine tune the last block along with the fully connected layers, the testing accuracy is 0.882. From Figure 3(d) we can see if more blocks are put for fine-tuning, the accuracy will increase, but the training time will also increase. The model reaches its limitation when fine tune the last three blocks.

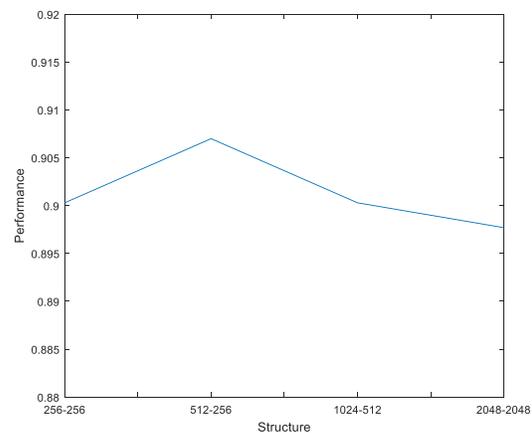

(a) Structure

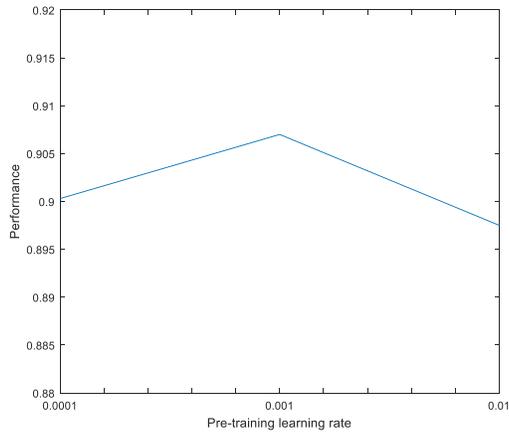
(b) Pre-training learning rate

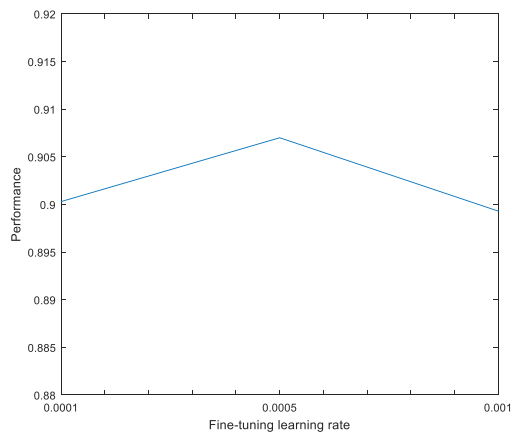
(c) Fine-tuning learning rate

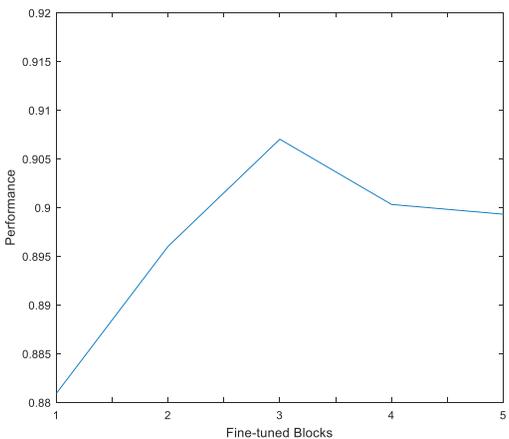
(b) Fine-tuning blocks
**FIGURE 3 Sensitivity Analysis of pre-trained deep CNN**

*Classification Result*

This section discusses the results with respect to RSC classification schemes under three different levels of granularity, namely, two classes, three classes and five classes. The performance of the classification results is measured using the following three indicators:
- Accuracy
- False positive rate
- False negative rate

*1) Two-class classification*

Figure 4(a) and 4(b) show the training and testing accuracy in each epoch for both pre-training and fine-tuning steps. In pre-training, the training accuracy starts at 56%, and reaches 75-77% after 50 epochs. The testing accuracy is generally higher than training performance, and it achieves over 80% accuracy after 50 epochs. In the fine-tuning step, we found that the training accuracy continues to increase from 76% to over 99% after 100 epochs, while the testing accuracy starts from 78%, and then levels off after 30 epochs, with the best performance of 90.72% (after 100 epochs)

False positive rate is another important measure of model performance, which is defined as the proportion of cases that road surfaces are in snow/ice presence conditions but are classified as bare pavement. From Table 2, for a bare surface, a false positive occurs when the model incorrectly classifies a bare surface as with snow covered, or vice versa. On the one hand, the implication of a high false positive for bare conditions is therefore a compromise in safety. On the other hand, the false positive of with snow covered detection has lower safety but higher risk of maintenance resource wastage due to false responses of incorrect classifications. Therefore, false positive for bare conditions is considered most critical. Some cases are shown in Table 2 as well.

*2) Three-class classification*

Figure 4(c) shows a comparison of the individual and overall RSC classification performance of alternative ANN, RT, RF, CNN and the Pre-trained deep CNN models proposed in this paper. Overall, the proposed model was successful at outperforming other machine learning models, showing 2.2%-4.4% improvement. For the validation data, bare surfaces were classified by the pre-trained deep CNN with 94.3% accuracy, compared to 91.8%, 93.8% and 93.8% by ANN, RT and RF respectively. For partly snow covered surfaces, the proposed model achieved 83.8% accuracy, while the other models only achieved accuracies between 75.1% and 77.4%. With an improvement of at least 8.27% in accuracy for this condition category, pre-trained deep CNN is successful in increasing the classification performance. However, for the classification of fully snow covered surfaces, the pre-trained deep CNN only performed better than the ANN model, having an accuracy of 56.3%; for this category, the best performance is the RF model with 71.2% accuracy. This limitation is likely caused by the small sample size of images for this particular class, which does not allow deep learning model to fully learn its features.

*3) Five-class classification*

The assessment conducted in the previous section was repeated using the more detailed five-class RSC description scheme. Figure 4(d) and Table 3 show the modeling statistics of the calibrated ANN, RT, RF, CNN and pre-trained deep CNN models. As can be seen in Table 3, the accuracy drops to 78.5%, lower than its counterpart that classifies RSCs according to the three-class terminology. This reduction is generally expected since there are two additional output classes being considered. The classification accuracy for bare, "<25", "25 to 50", "50 to 75" and ">75" snow covered surfaces as approximately 96.79%, 6.57%, 44.20%, 79.27% and 54.69%, respectively. This wide range in classification accuracy suggests the difficulty in distinguishing between certain RSC types (essentially bare and both wheel tracks bare) with the data available.

In this comparison, a three-layer ANN model consisting of an input layer, a hidden layer with 13 neurons and an output layer was calibrated with momentum of 0.1 and learning rate of 0.2. The final accuracy of the resulting ANN model is 75.6%. The RT model, which was constructed with four random variables at each node, was found to provide overall classification accuracy of approximately 77.9%, with prediction accuracy ranging from 28% to 94% for individual RSC classes. The RF model is comprised of 10 random trees with each tree being constructed while considering 4 random features out of a possible 14 at each node.

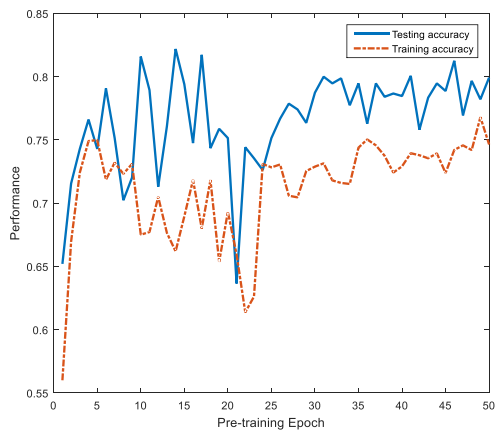

(a) Pre-training step in two-class scheme

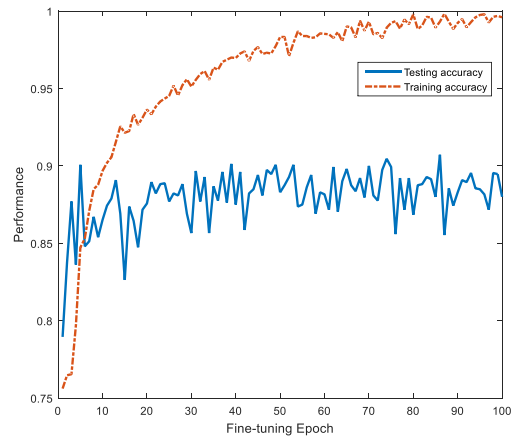

(b) Fine-tuning Step in two-class scheme

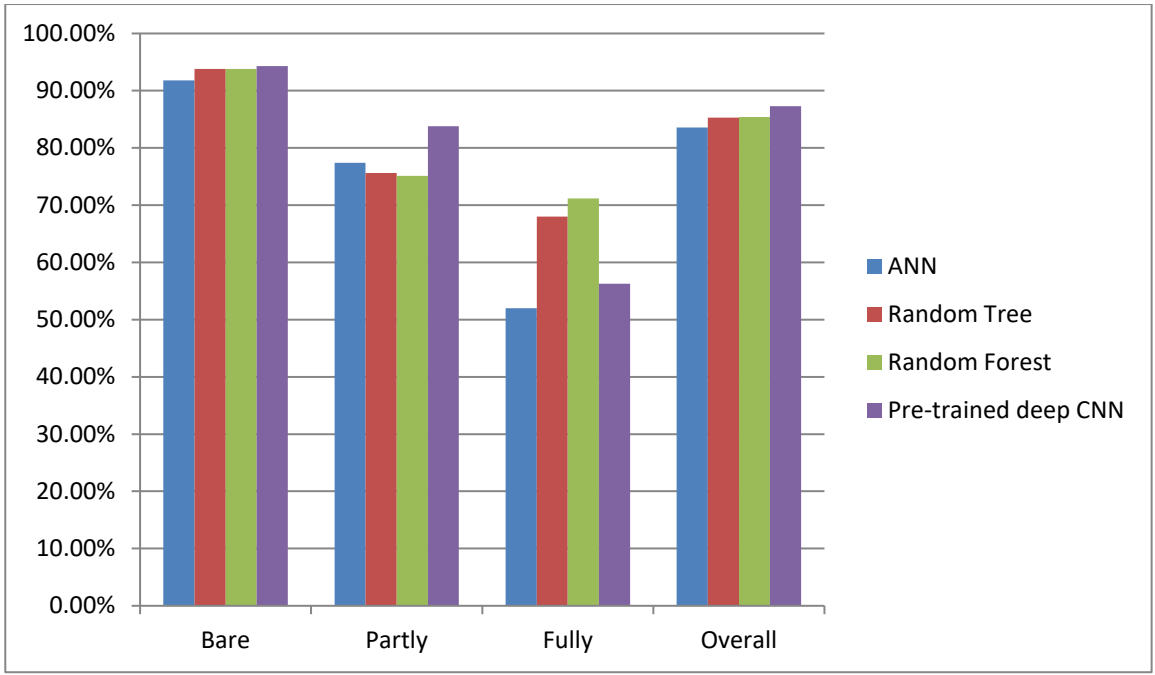

(c) Testing performances in three-class description

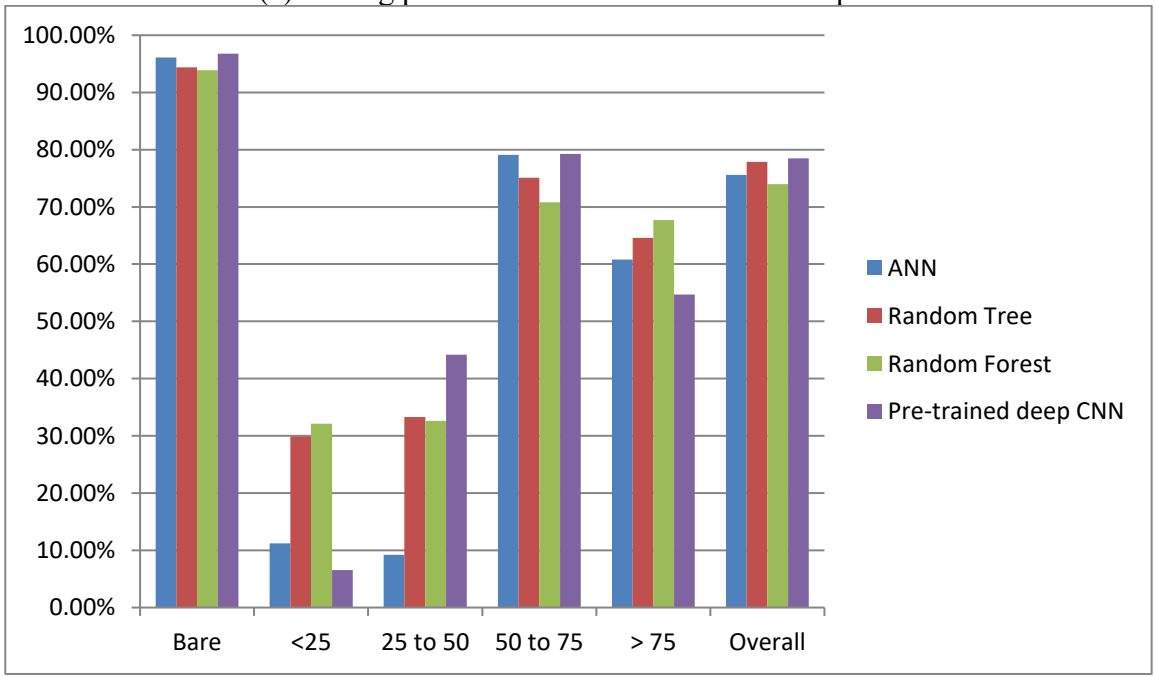

(d) Testing performances in five-class description

**FIGURE 4 Performance of the deep CNN model**

**TABLE 2 Two-class Description Testing Result**

| Ground Truth | False positive | Examples |
|---|---|---|
| With snow coverage 55.39% | 4.87% | 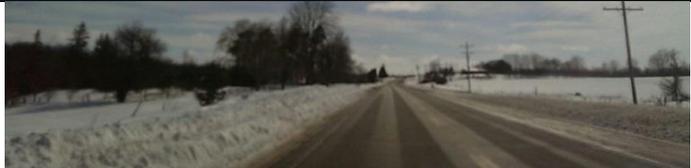 |
| Without snow coverage 44.61% | 14.75% | 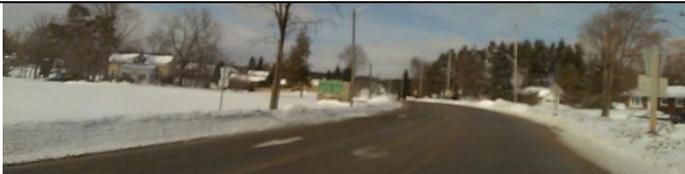 |
| Overall Accuracy | 90.72% | |

**TABLE 3 Performance Comparison of Models**

| Models | Two-Class Classification | Three-Class Classification | Five-Class Classification |
|---|---|---|---|
| ANN | / | 83.6% | 75.6% |
| Random Tree | / | 85.3% | 77.9% |
| Random Forrest | / | 85.4% | 74.0% |
| CNN | 88.4% | 84.8% | 76.7% |
| **Pre-trained CNN with localization** | **90.7%** | **87.3%** | **78.5%** |

Bare surfaces were classified extremely well by all five models, which showed accuracies between 94% and 96%, with the pre-trained deep CNN providing highest classification accuracy. On the other side, surfaces considered to be essentially bare (<25) were poorly classified by the proposed model, with an accuracy of less than 7%. Although still considered as poor performance, RT and RF demonstrated accuracies significantly greater at 30% and 32%, respectively. However, in practice it would not be uncommon for these images to be considered bare; therefore, if bare and "<25" conditions were combined into a single class, classification accuracy would dramatically increase up to approximately 84.16%, comparing to 85.37% and 85.25% by RT and RF. In conditions where the coverage is "25 to 50" and "50 to 75", pre-trained deep CNN still has the highest result, which shows better performance than the other models. Fully snow covered surfaces (>75) were classified with slightly lower accuracies than when using the three-class RSC description terminology. By observing the misclassification cases, we found that the majority of misclassifications were attributed to the "50 to 75" class, which may have similar maintenance decision-making implications. The RF outperformed all other models with approximately 68% classification accuracy.

*Performance vs. Data Size*

In this experiment, we further analyze the effect of data size on the performance of the proposed CNN models. The following bootstrapping process is followed:

1) Split the given dataset into two subsets: a training set and a testing set. The training set includes the 70% of data (3542 samples) while the testing set includes the remaining 30 % data (1521 samples).
2) A subset of data at a specific percentage is randomly drawn from the training data set (e.g., 10%, 50% and 100% of the total

training set) and then used to train and fine-tune the pre-trained deep CNN model. The trained model is subsequently used to classify the testing data set.
3) The training epochs are set to 50 and the fine-tuning epochs to 100. The training process will also stop when the training accuracy reaches 99%.

Figure 5 is a box plot of the test accuracy as a function of the training data size with the central mark representing the median, and the edges of the box for the 25th and 75th percentiles. As shown in the figure, the classification performance of the model is low at low data sizes, but increases quickly as the data size increases. The improvement trend of the model performance suggests that the deep learning model could reach a higher level of classification accuracy if there had been more training data available (e.g., it is not unrealistic to expect accuracy over 95% with sufficient amount of data). Furthermore, the variation of the model performance decreases as the data size increases, indicating the sensitivity of the model reliability with respect to data size. This result suggests that the performance of pre-trained deep CNN model improves as the data size increases.

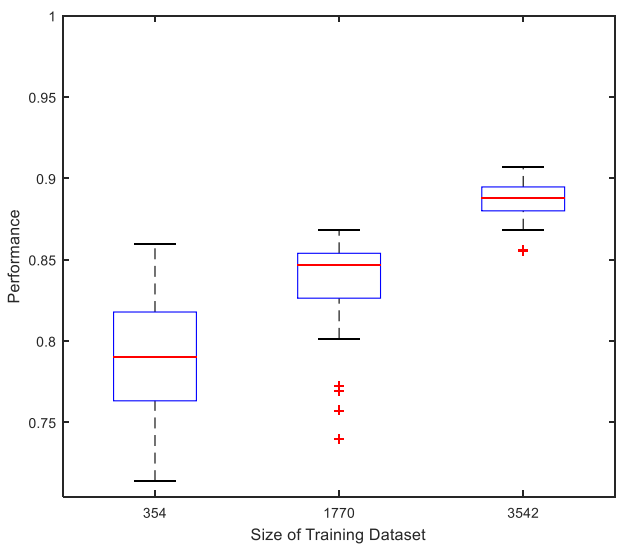

**FIGURE 5 Testing performances by different training data size**

## CONCLUSION

In this paper, we have investigated the potential of applying one of the most powerful machine learning techniques – deep learning, or deep neural networks, for classifying winter road surface conditions (RSC) based on image data. In particular, we proposed the idea of making use of a pre-trained convolutional neural network (CNN) with an addition of extra layers of neurons for model fine-tuning and localization. The resulting model is trained with our problem-specific training data – RSC image data. By analysing the model parameters and sensitivity, the best solution was found. The results have shown that the proposed idea of applying a pre-trained CNN model is effective in reducing the needs for large training data and computational time. The model was shown to outperform traditional machine learning models with performance advantage increasing as data size increases. The CNN model also has the advantage of being able to use the raw image data without pre-processing as required by most traditional approaches, facilitating its implementation and application.

This research represents an initial effort with several unsolved questions that need to be investigated in the future. For example, the robustness of the model performance to network structure needs to be further investigated so that the optimal network configuration for the problem of our interest could be determined. In this research, we adopted a pre-trained CNN model, which may reach its limit in classification capacity even it is trained using a large data set. Other types of pre-trained models and model structures such as Inception-3 (18) or deep residual learning model (22) should also be explored. The second issue is related to the need to adapt network structure to increased data size. Our preliminary analysis has shown the dependency of the performance advantage of a fixed model on data size. There may be further performance gain with adjusted model structure. Furthermore, a single type of images was used in this research; future research should explore the generality of the proposed method and model for a variety of images (e.g., those from dashboard camera and traffic cameras).

## ACKNOWLEDGEMENT

This research is supported by National Science and Engineering Research Council of Canada (NSERC). Authors would like to thank the Ministry of Transportation Ontario (MTO) for providing the data used in this study.